“THEME ARTICLE”, “FEATURE ARTICLE”, or “COLUMN” goes here: The theme topic or column/department name goes after the colon.

# Towards a QoE Model to Evaluate Holographic Augmented Reality Devices

## A HoloLens Case Study

**Longyu Zhang**
*University of Ottawa*
**Haiwei Dong**
*University of Ottawa*
**Abdulmotaleb El Saddik**
*University of Ottawa*

Augmented reality (AR) technology is developing fast and provides users with new ways to interact with the real-world surrounding environment. Although the performance of holographic AR multimedia devices can be measured with traditional quality-of-service (QoS) parameters, a quality-of-experience (QoE) model can better evaluate the device from the perspective of users. As there are currently no well-recognized models for measuring the QoE of a holographic AR multimedia device, we present a QoE framework and model it with a fuzzy-inference-system (FIS) to quantitatively evaluate the device.

The development of augmented reality (AR) technology has reshaped people's daily lives. AR merges virtual objects or information with the real-world environment to augment the reality and differs from virtual reality (VR) that blocks users from the real world. [1] Mobile AR games such as Pokemon Go have attracted thousands of players by placing virtual creatures on their smartphones, which simultaneously display the real-world surrounding environment. Rauschnabel et al. explored users' reactions to such mobile AR games from the aspects of hedonic, emotional and social benefits, social norms and physical risks.[2]

The current mobile phones, due to their limitations, cannot accurately reconstruct the surrounding environment to augment 3D contents precisely and vividly. Augmented reality smart glasses (ARSGs), such as Google Glass and Microsoft HoloLens, designed specifically for AR applications, provide users with a comparatively novel way to interact with AR contents.[3-4] Driven with various motivations and intentions, ARSGs have drawn increasing attention in several areas, such as medicine, tourism, education, social and marketing.[5] For instance, Ro et al. discussed the potential values and barriers of ARSGs;[3] Orts-Escolano et al. presented Holoportation, a system that allows the user to interact with the augmented avatars of remote users in real time;[6] while Kalantari et al. developed a model to understand individuals' acceptance of and reactions to ARSGs in a social environment.[7]

Compared to most other AR devices, HoloLens is able to operate independently as a complete AR system. In our previous paper,[8] we introduced the details of HoloLens' hardware components and conducted a series of technical evaluations with its functional components, which were primarily designed for performance-accuracy quality-of-service (QoS) testing. Compared to system-centric QoS evaluation methodologies, quality-of-experience (QoE) metrics are more human-centric and consider user-involved interactions, which is also of great value in providing a comprehensive overview of multimedia devices.

As QoE metrics vary depending on the system, to the best of our knowledge, there are no well-recognized QoE models for measuring ARSGs so far. Therefore, in this paper, we present a QoE framework containing three levels of influential parameters and model it with a fuzzy inference system (FIS) to quantitatively evaluate ARSGs. We validate our FIS model by comparing its outputs (the general user experience) with ground truth user ratings, also proving that the proposed QoE framework can represent a user's general AR experience. Details of the QoE framework and the FIS model are described, and validations and analysis of the model are also performed.

## QOE VERSUS QOS

Traditionally, aspects of system performance, such as network conditions and video quality, are usually evaluated with quality-of-service (QoS) models, proposed to capture the qualitatively or quantitatively defined performance contract between the service provider and the user applications. Although QoS models successfully measure the technological quality and functionality of systems with little human involvement, such models fail to evaluate the user's satisfaction with interactive multimedia applications or devices.[9]

Therefore, many research efforts have been undertaken to develop user-centric QoE evaluation models.[10] For instance, the European Network on Quality of Experience in Multimedia Systems and Services (QUALINET) has extended the notion of network-centric QoS to QoE in multimedia systems to develop subjective and objective quality metrics.[11] They distinguish one perception path and one reference path to explain the actual quality of the formation process and define QoE as “the degree of delight or annoyance of the user of an application or service”. Wu et al. also presented a conceptual framework of QoE in a distributed interactive multimedia environment and developed a mapping methodology to demonstrate the correlations between QoS and QoE.[12]

As various multimedia applications and devices are designed to satisfy users' varying needs, there are currently no unified QoE models applicable to all applications and devices. As a cutting-edge holographic AR device, the HoloLens is a representative device and can be used to create a corresponding QoE evaluation model.

## PROPOSED QOE EVALUATION FRAMEWORK

Based on the special attributes of AR devices, we present a conceptual QoE framework for holographic AR multimedia device evaluation. Our framework is primarily composed of four high-level (1st-level) parameters: content quality, hardware quality, environment understanding and user interaction. To clearly illustrate our proposed framework, we further explore each high-level parameter's lower-level (2nd- and 3rd-level) aspects to extend our framework. Figure 1 demonstrates the entire QoE framework for holographic AR multimedia device evaluation, with additional details provided in the following sections.

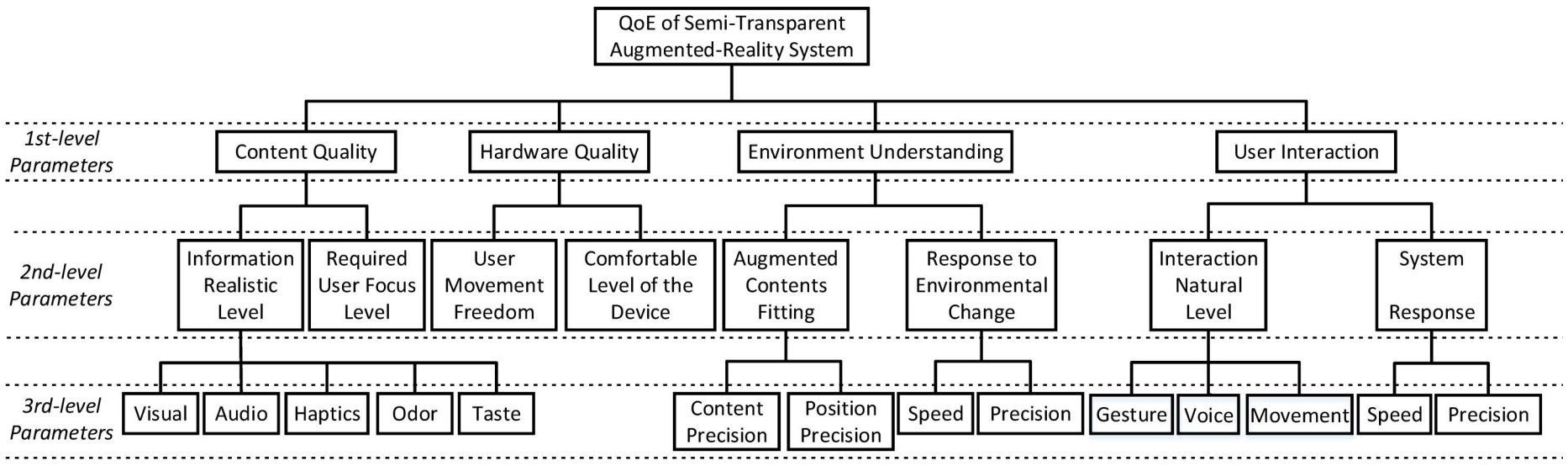

Figure 1. The proposed QoE framework for holographic AR device evaluation.

## Content Quality

Content quality is the core of an AR system. Two significant factors influencing the content quality are the information realistic level and the required user focus level.

First, the level of realism of contents generated by an AR device should at least meet the user's basic requirements. For instance, a 3D “cup” model, either created with graphics software or reconstructed from reality, should not look like a “book” when being augmented on a real table.[13] Similarly, a voice from the front should not sound as if its source is behind. Generally, AR usually has lower requirements regarding to the level of content realism than VR, which demands a vivid content to immerse users fully in the virtual world. In addition to the commonly mentioned visual and audio content, other properly integrated sensory content, such as haptic feedback, odor and taste, can also enhance the users' interaction experience.[14] Therefore, we also listed such content in our QoE evaluation framework; however, we did not put it in our HoloLens QoE user study questionnaire, as the selected HoloLens application does not provide such options.

Second, the required level of user focus varies with application and can also affect the required level of content realism. For instance, intense shooting games make users primarily focus on the target, instead of the surrounding environment. As a result, the realism level of the target should be higher than that of the background. Similarly, if augmented text can precisely deliver information to a user, its realism level is considered acceptable, regardless of the font, size or position.

## Hardware Quality

As users usually need to move around while wearing an AR device to interact with the augmented contents and the surrounding environment, AR hardware quality can greatly influence the user experience. The first concern should be with the user's freedom of movement. Traditional AR devices mostly require cables connected to external powerful computing devices to transmit and process the relevant content, largely restricting the movements of users. HoloLens solved this problem by embedding all computing parts and a Windows operating system inside the device itself. Another evaluation criterion is the level of comfort experienced while wearing an AR device. As most AR devices adopt the head-mounted display (HMD) design to enable prompt adjustments of the display screen's position based on the user's movements and direction of view, the user's head needs to support the entire device weight at all times. Therefore, the design and material quality would closely affect the user experience.

## Environment Understanding

As previously mentioned, compared to VR devices, AR devices have a special requirement to understand the real-world surrounding environment, which establishes the foundation before the reality is “augmented”. AR devices need to detect the locations of real-world objects around the user, estimate the user's relative position and the direction of view, and, occasionally, even recognize the objects or humans. Therefore, we consider environment understanding a unique aspect of a holographic AR device QoE evaluation model.

We divide this category into two sub-categories. First, the augmented elements need to fit the surrounding environment in both content and position to make it meaningful. Such a fit will allow the user to perceive a realistic and natural environment in the game. For instance, the generated virtual object on a real table should be a “cup” instead of a “car”; additionally, the “cup” should be at a proper location on the table instead of floating in the air. Second, as users change positions and direction of view frequently, AR applications should be able to respond to environmental changes quickly and precisely to avoid mismatches between real-world and augmented environments.

## User Interaction

In contrast to displaying images or playing movies, which merely shows content designed in advance, AR applications need to accommodate a certain amount of user interaction.[15] Thus, evaluating user interaction becomes an important part of our QoE framework. The first component is the natural level of interaction methods, which concerns the definitions of ways for users to interact with the application. Examples of such ways include waving a hand right or left to switch content, pressing a finger forward to push a “button” and saying “close” to stop the application. We summarize such defined interactions into commonly used methods, such as gesture, voice, and movement commands, in our framework. Another inevitable component is the precision and speed of system response to user interaction. Taking too long to respond to a user's interaction commands or misinterpreting certain commands can degrade a user's experience with holographic AR multimedia devices or even make users lose interest.

# FUZZY INFERENCE SYSTEM DESIGN

Theoretically, if the parameters in our proposed QoE framework accurately reflect users' experiences with a holographic AR device, we will be able to estimate users' general experience by analyzing users' ratings of such parameters. In other words, we can create a model that takes parameter ratings as inputs and generates the estimated general experience score as the output. Therefore, we perform a user study with HoloLens, asking participants to complete a questionnaire regarding all parameters in the framework (except haptic feedback, odor and taste aspects) and to provide a general rating score (a value between 0 and 100) for their AR experience. Subsequently, we use a part of the data to develop a fuzzy inference system (FIS) and test it with the remaining data. If the general scores estimated by the FIS model are similar to the general rating scores from users (the ground truth), then both our proposed framework and the FIS model are validated. Details of the user study will be provided in the next section discussing the experiment, while this section primarily introduces the design process of our FIS model.

## Selecting the Fuzzy Inference System

FIS maps the provided inputs to an output using fuzzy logic, similar to the process of human reasoning. It uses membership functions (MFs) to define the fuzziness in fuzzy set and computes output truth values using fuzzy rules. As QoE evaluation focuses particularly on the subjective feeling of the user and its parameters are also subjective and fuzzy in nature, being too complicated to be represented with classic linear approaches, we adopt the FIS to create our QoE evaluation model.

FIS primarily consists of four modules: a fuzzification module, a knowledge base, an inference engine and defuzzification. Mamdani FIS is well-known and commonly used.[16] Instead of directly using precise numbers or data as the input, Mamdani FIS performs fuzzification with membership functions representing degrees of membership. Subsequently, it builds fuzzy rules based on the acquired knowledge and utilizes the rule set to determine the output. As output is also fuzzy, a defuzzification method is required to calculate a crisp output value. This method is flexible with respect to adding or deleting rules.

Therefore, we define membership functions for inputs and the output and map the relationships between them by deriving fuzzy logic IF-THEN rules based on 60% of the user data. For instance, if we use a three-dimensional (3D) surface to represent the score mapping from two high-level parameters (content quality and environment understanding) to the overall rating score while setting the other high-level parameters to average values (50), we obtain their nonlinear relations as shown in Figure 2. The obtained general experience score calculated as the output of FIS may be compared to the user rating (the ground truth) to evaluate our FIS model.

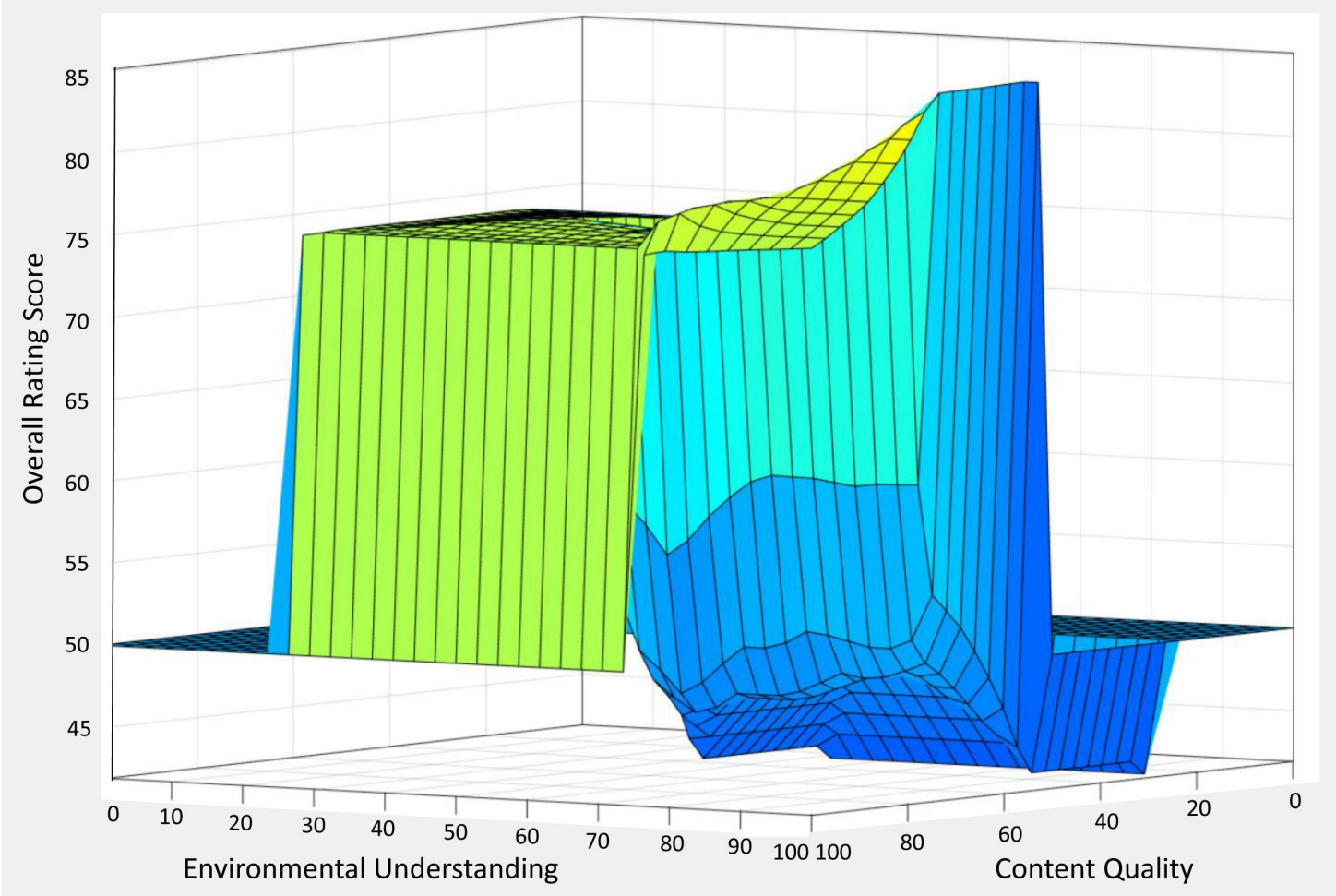

Figure 2. The 3D surface view showing the nonlinear mapping from content quality and environment understanding to the overall rating score.

## Choosing the Input and Output Variables

In the presented framework, we have introduced four high-level parameters and the corresponding low-level ones. To make the model more manageable and reduce the number of possible combinations, we chose the high-level parameters, including content quality, hardware quality, environment understanding and user interaction, as the input variables of the FIS. The model subsequently generates a crisp overall rating score ranging from 0 to 100 as the output variable.

To unify the measurements of input and output variables, we convert all input answers $I_{in}$, ranging from 1 to 5, into percentage values $I_p$, ranging from 0 to 100, as follows:

$$I_p = \frac{I_{in} - 1}{N - 1} \times 100 \tag{1}$$

where $N$ represents the number of answer options, being 5 in our case. Therefore, the raw input data are divided into five groups: 1→0, 2→25, 3→50, 4→75, and 5→100. In what follows, the input values are represented according to this percentage form.

## Defining Membership Functions

The fuzzy set theory considers degrees of truth and degrees of membership, while membership functions (MFs) represent the fuzzy subsets of each variable. Here, we first define MFs for variables of high-level parameters. As each question regarding high-level parameters in our designed questionnaire has 5 answers (1 to 5) to choose from, we divide each input variable into 5 equal MFs with triangular shapes, namely, very poor, poor, fair, good and excellent. Each input value is then transformed to the corresponding MF based on its degree of membership. For instance, the value of 50 is “Fair”, while the value of 75 is “Good”. Our user study results show that no rating scores are in the “very poor” group, which was thus omitted.

As the overall rating score is more diffuse than the input scores, we use the fuzzy c-means (FCM) clustering method to define the output's fuzzy sets before generating MFs, similar to the approach of Hamam et al.[10] As all scores from the questionnaires are greater than 50, we divide the output into 4 groups. The detailed results are provided in the next section.

## Deriving Fuzzy Rules

As we have defined MFs for all high-level parameters and the overall rating score, user data from questionnaires can hence be transferred to the corresponding MFs to build fuzzy rules. For instance, if a user rates content quality, hardware quality, environment understanding, user interaction and the overall rating as 75, 50, 100, 75, and 85, respectively, these values can subsequently be interpreted into MFs as Good, Fair, Excellent, Good, and Good, respectively, and a fuzzy rule can be derived as follows:

*IF* content quality is *Good*,

*AND* hardware quality is *Fair*,

AND environment understanding is *Excellent*,

*AND* user interaction is *Good*,

*THEN* the overall rating is *Good*.

## Generating Output

Given a set of inputs, based on the fuzzy rules, the FIS model can generate the corresponding output. As our system is a Mamdani-type FIS, a defuzzification process for the results is required to obtain a crisp value as the output.

Defuzzification is the process of generating a quantifiable result in crisp logic using the given fuzzy sets and the corresponding membership degrees. It maps a fuzzy set to a crisp set. In our implementation, we used the centroid calculation to calculate the center of gravity of the curve describing the output. Therefore, the FIS output becomes a crisp value representing an overall rating score of our presented model.

# EXPERIMENT AND RESULTS

During the user study, users were first trained to become familiar with the HoloLens, subsequently used it to play one of two selected AR applications, and finally, completed} the questionnaire. Details of the HoloLens application selection, questionnaire design, experimental setup, results and analysis are as follows.

## HoloLens Application Selection

The AR applications we used for testing are a first-person shooting game “RoboRaid” (Figure 3) and an adventure game “Young Conker” from Microsoft. We chose these two applications because they both cover all 1st-, 2nd-, and 3rd-level parameters (except for haptic feedback, odor and taste aspects) of the framework. The selected ARSG applications differ from those of mobile AR games (e.g., Pokemon Go) in several ways, such as 3D holographic displays, enabling more natural gesture control and augmenting content based on the detailed surrounding environment instead of directly superimposing content on the screen. Participants, to better experience these special attributes of ARSGs during the test, were required to perform the following related actions (we use “RR” to denote “RoboRaid” and “YC” to denote “Young Conker”):

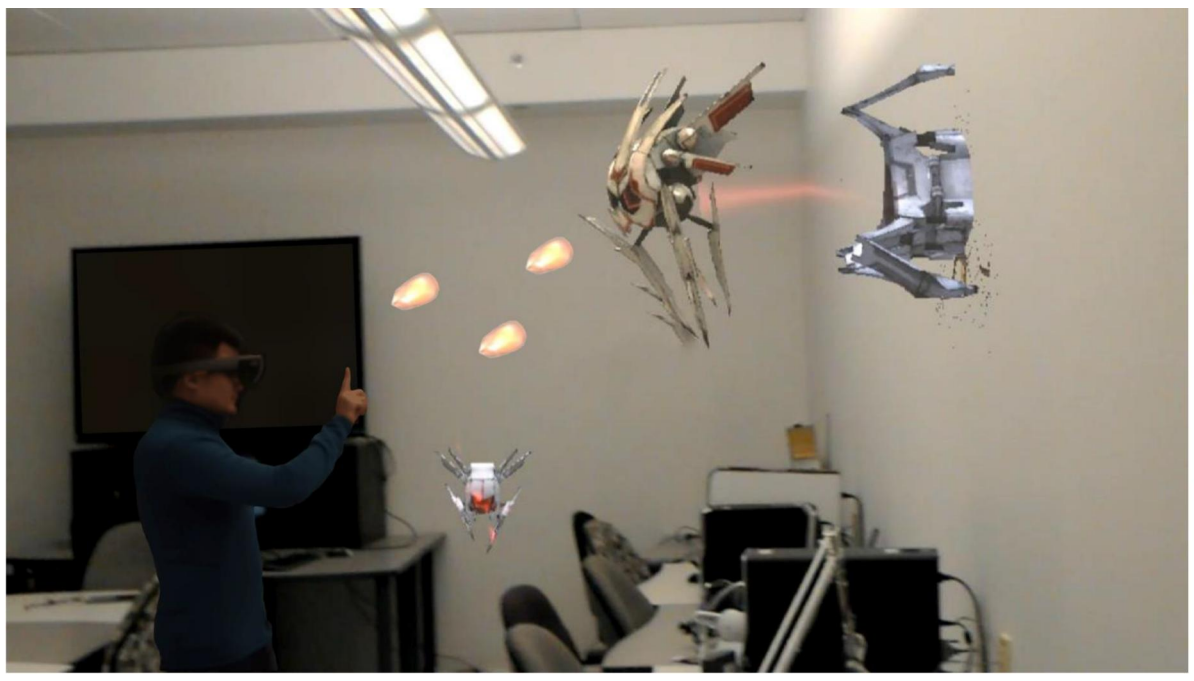

(a) A user is playing *“RoboRaid”*

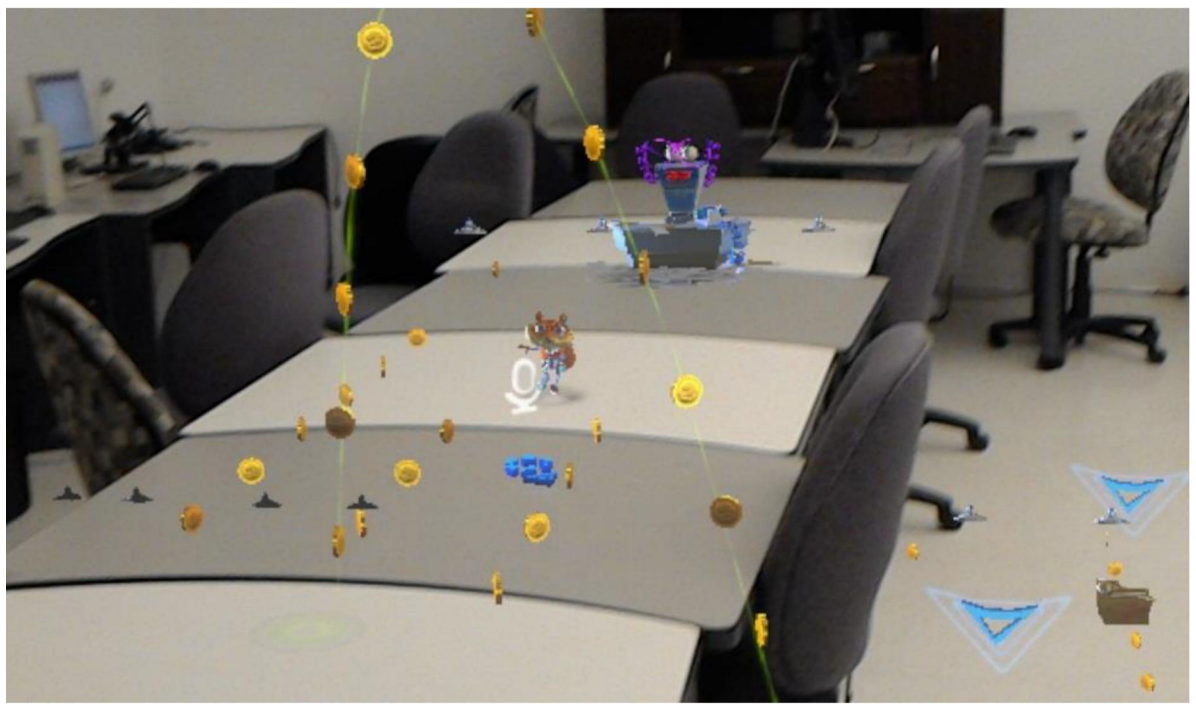

(b) A user is playing *“Young Conker”*

Figure 3. The scenarios of users playing with AR applications “RoboRaid” and “Young Conker”.

- Content quality: for “RR”, watch the augmented enemies and their fire and listen to the noise direction to decide which part of the real-world wall would be “breached” by battleships; for “YC”, browse coins, chase or avoid enemies, follow commands of NPCs, etc.
- Hardware quality: both “RR” and “YC” need users to walk and run around the room wearing HoloLens to adjust the distance from enemies.
- Environment understanding: “RR” needs users to find enemies “climbing” on the wall or “flying” in the air and to “destroy” the real-world wall to reveal the hidden aliens; “YC” requires participants to control the avatar to jump on a real-world desk or virtual coins with a springboard on the actual ground.
- User interaction: “RR” uses an air-tap to shoot, moving body to dodge attacks, and saying “X-Ray” to activate the special see-through weapon; “YC” defines the gazing direction as the avatar's direction of movement, saying “let's go” to start, etc.

## Questionnaire Design

A questionnaire was designed for the user study based on our proposed framework, and related documents are posted on GitHub (github.com/HoloLensQoE/HoloLens_QoE_User_Study). The questions for each application are identical, except for the examples to help participants understand questions specifically selected based on the game. The questionnaire first collects several aspects of basic user information, including name, gender, age range and previous AR and non-contact gesture control experience. For each related QoE parameter, we design a Likert-scale question with the answer provided on a five-point Likert scale, namely, with values from "1" to "5" representing different responses (e.g., from "not at all" to "completely"). Questions regarding low-level parameters are asked first before a general rating for the corresponding high-level parameters is required. For instance, Question 1 (Q1) asks the participant to rate the realism level of visual information on a scale from 1 to 5, namely, from "not at all" to "completely realistic"; Q2 concerns the realism level of audio, while Q3 asks for the required focus or attention level. After these three questions, a general question G1 is presented to rate the high-level parameter content quality based on Q1-Q3. As a result, the participant would have a better understanding of high-level parameters. Finally, a rating score (0-100) for the overall quality of this AR experience is also required.

## Experimental Setup

The user study was performed in a 14.5 m × 9.5 m room. A total of 75 subjects (47 male and 28 female) participated in this test, of whom 8 were in the age range of 10-19, 44 were aged 20-29, 20 were aged 30-39, and 3 were aged 40-49. In addition, 28 users never used AR-related applications, while 23 users had no experience with non-contact gesture control applications. Sixty users tested the application “RoboRaid”, while 15 users played “Young Conker”.

Participants were given 10 minutes to become familiar with HoloLens by performing interactions, such as moving the head to aim the menu, displaying and dismissing the main menu, using the air-tap gesture or the “select” voice command to confirm selection, and making the pinch gesture to move or manipulate the target. After participants became familiar with HoloLens, we provided more detailed instructions for the selected application and performed the test. Finally, a questionnaire, designed based on our proposed QoE evaluation framework, was completed by the participants. We also used a laptop to mirror the views of the users in real time through the HoloLens device portal API to provide users with additional instructions when needed. A Lenovo Y50 laptop was used, with a 15.6" screen with 1920×1080 resolution, an Intel i7-4710HQ CPU, 16 GB of RAM, and a Windows 10 64-bit operating system.

## FIS Model Results

The general architecture of our implemented FIS model is shown in Figure 4, which contains MFs defining results for four inputs (content quality, hardware quality, environment understanding and user interaction) and one output (the overall rating score), as well as the derived 43 fuzzy rules.

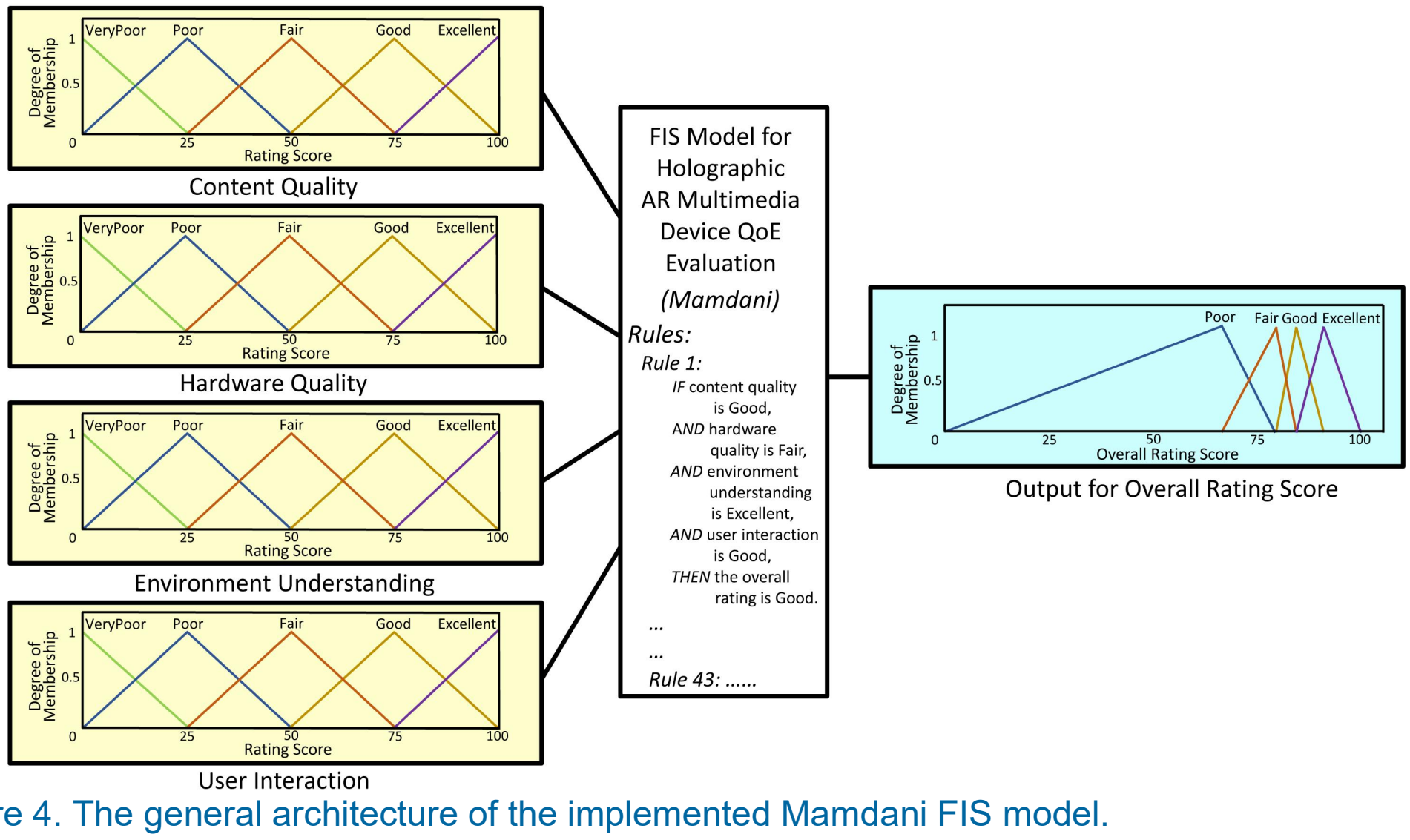

Figure 4. The general architecture of the implemented Mamdani FIS model.

## MFs for Input and Output

As previously introduced, each input variable is equally divided into 5 MFs with triangular shapes corresponding to the 5 answer choices (Figure 4 inputs). As the output variable is more diffuse, we first apply the FCM clustering method and subsequently define its MFs.

The FCM divides the overall rating scores into different clusters based on each value's degree of belonging to various groups. As Figure 5 shows, each value is represented by the shape and color of its defined cluster (purple '*', green 'o', blue '+' and red '×'), while the centers of all clusters are displayed with larger black symbols, with rating values of 90.80, 84.41, 79.58 and 66.58. These values are subsequently used as the peak points of the MFs of the FIS output accordingly (Figure 4 output).

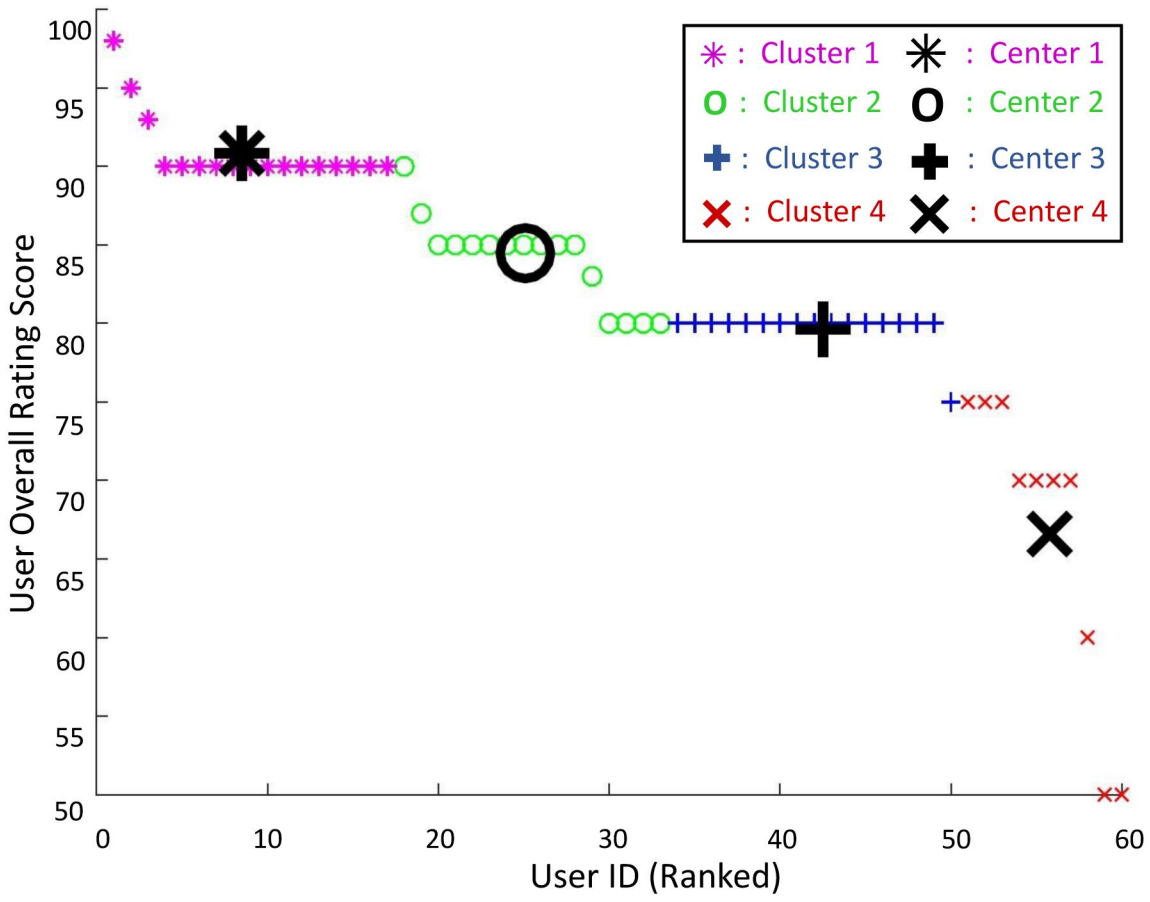

Figure 5. Fuzzy c-means (FCM) clustering results for the user overall rating score.

## Fuzzy Rules Derived

As the data of 45 "RoboRaid" users (60% of all user data) are used to derive the fuzzy rules and the data from each questionnaire can be used to build one rule, a few rules may be repeated or in conflict. If a rule occurs more than once, we multiply the base weighting value by such rule's frequency of appearance to obtain its new weighting value; we add conflicting rules as separate fuzzy rules to balance the result. Based on this strategy, we created 43 fuzzy rules for our FIS model (Figure 4 rules).

# Statistical Analysis

After building the FIS model, we consider the data of the remaining 15 "RoboRaid" users and 15 "Young Conker" users as the testing data. Each subject's data outputs have a ground truth overall user rating $QoE_u$ from the user and an FIS-estimated-rating $QoE_f$ from our model. The testing results are shown in Table 1. We first explore the descriptive statistics, including the mean, 95% confidence interval, standard error, median, and standard deviation, of each series of data individually to form the first impression; subsequently, we compare each application's root-mean-square errors (RMSEs) of $QoE_u$ and $QoE_f$. Finally, we compare them and perform a paired-samples T-test to statistically analyze them.

Table 1. Testing results of the overall user rating ($QoE_u$) and the FIS-estimated rating ($QoE_f$)

| Testing Application | QoE | Mean | 95% Confidence Interval for the | Std. | Median | Std. | RMSE | Paired T- |
|---|---|---|---|---|---|---|---|---|

| | | | mean | | Error | | Deviation | (%) | Test |
|---|---|---|---|---|---|---|---|---|---|
| | | | Lower Bound | Upper Bound | | | | | P-Value |
| RoboRaid | $QoE_u$ | 85.133 | 82.776 | 87.491 | 1.099 | 85.0 | 4.257 | 3.895 | 0.493 |
| | $QoE_f$ | 85.853 | 83.450 | 88.257 | 1.121 | 85.0 | 4.340 | | |
| Young Conker | $QoE_u$ | 86.867 | 82.282 | 91.452 | 2.138 | 90.0 | 8.280 | 5.791 | 0.728 |
| | $QoE_f$ | 87.413 | 84.427 | 90.399 | 1.392 | 91.9 | 5.392 | | |

## Descriptive Statistics

Descriptive statistics for each individual overall user rating $QoE_u$ and FIS-estimated rating $QoE_f$ are shown in Table 1 for initial analysis. We observe that each pair of $QoE_u$ and $QoE_f$ for a given application are similar. For instance, the “RR” value of $QoE_u$ has an average of 85.133 with a 95% confidence interval of 82.776 to 87.491, a standard error of the mean ($SE_M$) of 1.099, a median of 85.000, and a standard deviation of 4.257; the corresponding “RR” value of $QoE_f$ has a mean of 85.853 with a 95% confidence interval of 83.450 to 88.257, an $SE_M$ of 1.121, a median of 85.000, and a standard deviation of 4.340. Figure 6 shows the $QoE_u$ and $QoE_f$ for “RoboRaid” and “Young Conker”, respectively. From this figure, we can see that each pair of $QoE_u$ and $QoE_f$ has similar patterns.

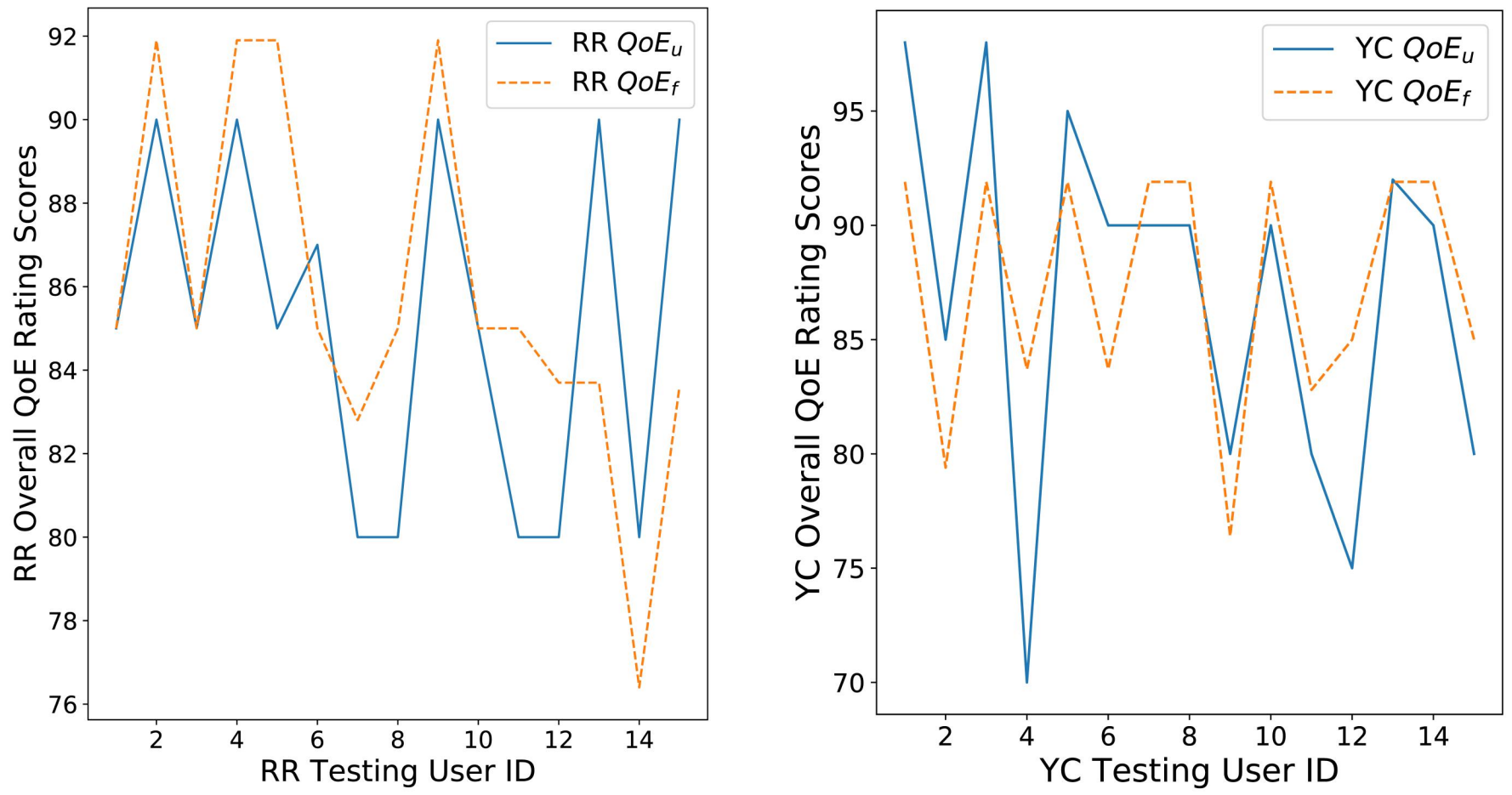

Figure 6. Overall QoE rating scores by users ($QoE_u$) and FIS model ($QoE_f$) for “RoboRaid” (RR) and “Young Conker” (YC), respectively.

## Root-Mean-Square Error

Besides comparing the descriptive statistics of user data, we also further evaluate the differences between each pair of $QoE_u$ and $QoE_f$ through computing their root-mean-square errors (RMSEs), which is the square root of the average of squared errors. As shown in the Table, the RMSE for “RR” is 3.895, indicating that, on average, our FIS-estimated rating deviates from the ground-truth value of the user rating by 3.895 points on a scale of 100. Similarly, we obtain an RMSE for “YC” of 5.791 on a scale of 100.

### Paired-Samples T-Test

To determine if the difference between each pair of $QoE_u$ and $QoE_f$ is statistically significant, we performed the paired-samples t-test and compared the p-value to the standard significance level $\alpha$=0.05. Our null hypothesis can be expressed as $H_0$: $\mu_1 = \mu_2$, namely, there is no significant between the two series of values. If the p-value is smaller than or equal to $\alpha$, we can reject the null hypothesis and conclude that the difference between the pair is statistically significant. Otherwise, the null hypothesis holds. The table shows p-values for "RR" and "YC" of 0.493 and 0.728, respectively, with both being larger than the significance level $\alpha$ (0.05). Therefore, the null hypothesis $H_0$ holds, and there are no statistically significant differences between each pair of $QoE_u$ and $QoE_f$, i.e., our FIS-estimated ratings are not significantly different from the ground truth overall user ratings.

## DISCUSSION

During our user study, participants reported several advantages and disadvantages of HoloLens affecting their experience. Thus, we compare user feedback to the questionnaire results to discuss their correlation.

First, most participants enjoyed the freedom of movement offered by HoloLens due to its capability to process all data by itself without the requirement of connecting to an external computing device with a cable (average score 3.92 out of 5). However, as HoloLens is a completely integrated head-mounted display (HMD) device, its heavy weight degrades the user's comfort level while wearing it and causes fatigue after a while. Thus, the average score of device comfort is merely 3.21. As a result, the users' high-level parameter hardware quality obtains a score of 3.63.

Second, though HoloLens can display augmented 3D contents upon the real-world surrounding environment vividly with an advanced holographic projection technology, several participants complained about its limited field of vision, affecting the visual quality score (average 3.83).

Third, as gesture commands are detected and recognized with a small-size depth camera embedded inside the front part of the HoloLens, it can recognize only a few gestures within a small region. Therefore, the air-tap gesture, instead of more natural gestures (e.g., the finger gun gesture), is defined as the shooting action in the "RoboRaid" application, obtaining a score of only 3.16. On the other hand, using body movement to dodge attacks provides participants with a natural interaction experience; the average score is 4.24.

These findings also prove that the parameters in our proposed QoE framework for holographic AR multimedia device evaluation can represent participants' actual experiences and would be helpful for the design of such devices and applications with specific aspects.

## CONCLUSION

In future work, we plan to integrate haptic feedback, digital odor, and artificial taste with HoloLens to explore their influence on users' QoE, which would provide additional experience beyond audio and video. Measuring and evaluating these multimodal attributes for holographic AR multimedia device would be of great value to guide developers and designers to improve their products.

## REFERENCES

1. M. Billinghurst et al. A survey of augmented reality. Foundations and Trends in Human–Computer Interaction, 8:73–272, 2015.
2. P. Rauschnabel et al. An adoption framework for mobile augmented reality games: The case of pokémon go. Computers in Human Behavior, 76:276–286, 2017.

3. Y. Ro et al. Augmented reality smart glasses: Definition, concepts and impact on firm value creation. In Augmented Reality and Virtual Reality, pages 169–181. Springer, 2018.
4. Z. Lv et al. Hand-free motion interaction on Google Glass. In proceedings of SIGGRAPH Asia 2014 Mobile Graphics and Interactive Applications, page 21, 2014.
5. P. Rauschnabel. Virtually enhancing the real world with holograms: An exploration of expected gratifications of using augmented reality smart glasses. Psychology & Marketing, 2018.
6. S. Orts-Escolano et al. Holoportation: Virtual 3D teleportation in realtime. In Proceedings of the 29th Annual Symposium on User Interface Software and Technology, pages 741–754, 2016.
7. M. Kalantari and P. Rauschnabel. Exploring the early adopters of augmented reality smart glasses: The case of Microsoft Hololens. In Augmented Reality and Virtual Reality, pages 229–245. 2018.
8. Y. Liu et al. Technical evaluation of Hololens for multimedia: A first look. IEEE Multimedia, 2018.
9. R. Jain et al. Quality of experience. IEEE MultiMedia, 11:95–96, 2004.
10. A. Hamam et al. A quality of experience model for haptic virtual environments. ACM Transactions on Multimedia Computing, Communications, and Applications, 10:28:1–28:23, 2014.
11. P. Le Callet et al. Qualinet white paper on definitions of quality of experience. European Network on Quality of Experience in Multimedia Systems and Services, 3:1–19, 2012.
12. W. Wu et al. Quality of experience in distributed interactive multimedia environments: toward a theoretical framework. In proceedings of the 17th ACM International Conference on Multimedia, pages 481–490, 2009.
13. L. Zhang et al. From 3D sensing to printing: A survey. ACM Transactions on Multimedia Computing, Communications, and Applications, 12:27:1–27:23, 2015.
14. A. El Saddik et al. Haptics technologies: Bringing touch to multimedia. Springer Science & Business Media, 2011.
15. L. Zhang et al. Visualizing Toronto city data with HoloLens: Using augmented reality for a city model. IEEE Consumer Electronics Magazine, 7:73–80, 2018.
16. A. Hamam et al. User force profile of repetitive haptic tasks inducing fatigue. In proceedings of the 7th International Workshop on Quality of Multimedia Experience, pages 1–6, 2015.

# ABOUT THE AUTHORS

**Longyu Zhang** is currently pursuing his Ph.D degree in Electrical and Computer Engineering at Multimedia Communications Research Laboratory (MCRLab), University of Ottawa. He has been awarded Canada NSERC Postgraduate Scholarship (Doctoral) for 2015-2018. His research interests include augmented reality, computer vision, and multimedia. Contact him at lzhan121@uottawa.ca.

**Haiwei Dong** is a Research Scientist at the School of Electrical Engineering and Computer Science at the University of Ottawa. His research interests include robotics, multimedia, and artificial intelligence. Dong received his Ph.D. from the Graduate School of Engineering at Kobe University. He is a Senior Member of IEEE and a licensed Professional Engineer. Contact him at hdong@uottawa.ca.

**Abdulmotaleb El Saddik** is a Distinguished University Professor at the University of Ottawa. His research focus is on multimodal interactions with sensory information in smart cities. He is an ACM Distinguished Scientist, a Fellow of the Engineering Institute of Canada, a Fellow of the Canadian Academy of Engineers and an IEEE Fellow. He received the IEEE I&M Technical Achievement Award and the IEEE Canada Computer Medal. Contact him at elsaddik@uottawa.ca.